\documentclass[aps,twocolumn,groupedaddress,showpacs]{revtex4}

\usepackage[dvips]{graphicx}
\usepackage{epsfig}

\begin{document}

\title{Effect of interactions on the noise of chiral Luttinger liquid systems}

\author{B.~Trauzettel$^{1,2}$, P.~Roche$^3$, D.~C.~Glattli$^{3,4}$, and 
H.~Saleur$^{2,5,6}$}

\affiliation{$^1$Laboratoire de Physique des Solides, Universit{\'e} 
Paris-Sud, 91405 Orsay, France\\
$^2$Physikalisches Institut, Albert-Ludwigs-Universit\"at, 79104 Freiburg, 
Germany\\
$^3$Service de Physique de l'Etat Condens{\'e}, CEA Saclay, 91191 
Gif-sur-Yvette, France\\
$^4$Laboratoire Pierre Aigrain, D{\'e}partement de Physique de l'Ecole Normale 
Sup{\'e}rieure, 75231 Paris, France\\
$^5$Department of Physics, University of Southern California, 
Los Angeles, CA 90089-0484, USA\\
$^6$Service de Physique Th{\'e}orique, CEA Saclay, 91191 
Gif-sur-Yvette, France}

\date{\today}

\begin{abstract}

We analyze the current noise, generated at a quantum point contact in 
fractional quantum Hall edge state devices, using  the chiral 
Luttinger liquid model with an impurity and the associated exact field 
theoretic solution. We demonstrate that an experimentally 
relevant regime of parameters exists where the noise coincides with the 
partition noise of independent Laughlin quasiparticles. However, outside of 
this regime, this independent particle picture breaks down and the inclusion
of interaction effects is essential to understand the shot noise.

\end{abstract}


\maketitle

In fractional quantum Hall (FQH) devices at filling factor $\nu=1/3$ 
the effect of electron interactions turns the bulk of the system 
into an incompressible quantum fluid. The edge excitations are gapless, 
and can be described  by the chiral Luttinger liquid 
(CLL) model \cite{wen90}. This model exhibits quasiparticles (QPS) which can be thought 
of as collective excitations of the electronic system. One of the consequences 
of strong electron interactions is that these QPS have fractional charge 
$e^*=e\nu$ \cite{laugh83}. In many other respects however, these QPS do 
behave like free electrons; for instance, in the absence
of a dominant impurity, i.~e. a quantum point contact, the right 
and left moving Laughlin QPS move  without 
scattering along the edges of the sample. 
However, once a quantum point contact brings the edges close together, 
QPS can tunnel from one edge to the other
near the constriction. In this tunnelling process, they do not 
behave like independent particles any longer: 
the tunnelling term has a very confining action,
and can, in the strong backscattering limit (SBL), bind them
into triplets with the charge of an electron.
This difference to the naive picture of non--interacting Laughlin QPS in the
bulk of a FQH device  has led in our opinion to 
confusing interpretations of transport experiments for
FQH edge states. A particularly elaborated example is the measurement
of fractional charge through shot noise 
\cite{depic97,samin97,glatt00,griff00,chung03}. Most of these 
measurements
have been successfully described by essentially non--interacting formulae, 
where  the effect of electron interactions is 
usually represented by  the fractional charge, and
sometimes by an energy--dependent transmission coefficient.
From the CLL point of view, however, it seems to be clear that the tunnelling 
process involves interaction effects which should, in general, not 
be encompassed by independent particle formulae.

In view of these problems, it has even been suggested that the CLL model
is not sufficient to explain the measurements of $e^*=e/3$ and
more complex mechanisms need to be taken into account. \cite{rosen01}
In order to assess the relevance of such mechanisms it is necessary to study 
further the physical predictions for the noise in tunnelling experiments 
coming from the model of a CLL with an impurity. This is a nontrivial 
exercise, as the problem is highly non--perturbative. 
Although an exact solution exists, this solution is complex, and its
physical implications had not been much explored. We revisit this solution in the 
present paper, and find, surprisingly maybe, that a rather wide regime of parameters 
{\sl does} exist, in which Laughlin QPS behave as if they were 
indeed tunnelling independently. 
There is, however, a complementary  regime of parameters, in particular, 
at low temperatures, where the effect of interactions is very 
pronounced and cannot be incorporated 
in an independent particle model over an extended range of the
applied voltage anymore. 
The latter regime is probably as relevant as the former for experiments, 
and we expect our work to spur further analysis of data deviating from  
non--interacting approximations.

The CLL model is described by the Hamiltonian
\begin{equation}
H_0= \frac{\hbar \pi v}{g} \int dx \left( n_R^2 + n_L^2 \right) \; ,
\end{equation}
where the  right and left moving densities are related to
Bosonic phase fields $\phi_{R/L}$ by $n_{R/L} = \pm \partial_x
\phi_{R/L}/2\pi$, $v$ is the plasmon velocity, and $g$ the
standard Luttinger interaction parameter (see e.g.~\cite{fishe97}).
It has been shown that the right and left modes $n_{R/L}$ are
mathematically isomorphic to the right and left moving edge states
in a FQH bar with filling fraction $\nu=1/m$ ($m$ odd integer) 
\cite{wen90} provided one takes $g=\nu$. An impurity at site $x=0$ 
in such a system corresponds to adding the dominant $2k_F$
backscattering contribution
\begin{equation} \label{HB}
H_B = \lambda \cos \left[ \phi_L(0)-\phi_R(0) \right]
\end{equation}
to the Hamiltonian. This term hops a quasiparticle $e^{i\phi_{R/L}}$, 
i.~e. a right/left mover, between the two modes. The charge
transfered in such a hopping process is fractional and given by
$e^*=eg$ \cite{fishe97}. From the latter two Hamiltonians it
becomes clear that the right and left moving edge states in FQH
bars are non--interacting in the bulk, but
do interact with each other at the quantum point contact. A symmetrically
applied external bias $V$ can be modelled by a third term of the
Hamiltonian $H_V = (eV/2) \int dx ( n_L - n_R)$.

We are particularly interested in the effect of interactions on
the zero frequency current noise
\begin{equation} \label{def_noise}
S = \lim_{\omega \rightarrow 0} \int dt e^{i \omega t} \left\langle 
\left\{ \Delta I (t) , \Delta I (0) \right\}_+ \right\rangle
\end{equation}
with the current fluctuation operator $\Delta I(t) = I(t) - \langle I
\rangle$. If uncorrelated electrons in a single channel quantum wire, 
connected to reservoirs with a voltage bias $V$, are partly 
backscattered off an impurity in the wire, the voltage and temperature
dependence of the current noise in a two--terminal setup can be
written as \cite{lesov89}
\begin{equation} \label{leso1}
S = 2 G_0 {\cal T}(1-{\cal T}) e V \coth
\left(\frac{e V}{2k_B T} \right) + 4 k_B T G_0 {\cal T}^2 \; ,
\end{equation}
where ${\cal T}$ is the energy--independent transmission coefficient of 
the impurity, $G_0 = e^2/h$, and $T$ the temperature.
The simplest question one can ask is whether a formula similar to 
Eq.~(\ref{leso1}) could describe (maybe with an appropriate carrier charge)
the noise due to tunnelling of Laughlin QPS 
through a constriction in a FQH bar. This has been assumed 
in the interpretation of some shot noise experiments 
\cite{depic97,griff00} with apparent success. 

A first remark is in order. Simple Keldysh calculations in the CLL with 
an impurity give for the shot noise \cite{kane94}
\begin{equation} \label{kfwbl}
S=2 e^* \coth\left( \frac{e^* V}{2k_B T} \right) I_{\rm BS}
\end{equation}
in the weak backscattering limit (WBL) where $e^*=e g$ and 
$I_{\rm BS}$ is the backscattered current. Likewise,
\begin{equation} \label{kfsbl}
S=2e \coth \left( \frac{eV}{2k_B T} \right) I_T
\end{equation}
in the SBL where $I_T$ is the transmitted current.
The former result suggests the picture of uncorrelated Laughlin QPS 
tunnelling from one edge to the other at the constriction. The latter 
result can be similarly interpreted 
as the uncorrelated tunnelling of electrons through the constriction. 
It should therefore be clear qualitatively that 
the tunnelling induces strong interactions between the Laughlin QPS which get  
essentially ``confined'' in the SBL. 
This might appear puzzling at first sight. After all, the tunnelling term has 
the same form as the term one would write for independent electrons 
$H_B\propto \psi^\dagger_R\psi_L+ \; {\rm h.c.} \; $. The subtlety is that 
the QPS operators 
$\psi_{R/L} = e^{\pm i\phi_{R/L}}$ do not behave at all like usual 
creation/annihilation operators. 
For instance, acting on a state, they do more than just adding or annihilating
Laughlin QPS, 
but they modify the whole ``sea'' of them, because of their nontrivial 
commutation 
relations. Therefore, processes involving several Laughlin QPS are not 
reducible in 
any simple way to single particle processes, and an interaction between 
Laughlin QPS 
is induced through the tunnelling.
Another way to see this is to observe that under renormalization, 
$H_B$ is not invariant, 
but renormalizes. Thus, the balance 
of multiparticle processes depends on the energy scale, a typical result for 
problems with strong interactions.

It is clear experimentally that  Eqs.~(\ref{kfwbl}) and
(\ref{kfsbl}) have a limited range of validity,
and cannot  fit much of the existing  data
\cite{depic97,samin97,glatt00,griff00,chung03}.
Non--perturbative results for the current noise are thus  needed to
compare experimental observations with CLL theory. The
exact field theoretic solution of the CLL model with an impurity,
carried out in Ref.~\cite{fendl95}, allows for a non--perturbative
calculation of the current noise, even at finite temperature \cite{fendl96}. 
The main steps of this calculation are as follows. 
A CLL with an impurity in the presence of a DC bias,
defined by the Hamiltonian $H=H_0+H_B+H_V$, can be mapped onto the 
boundary sine--Gordon model \cite{fendl95}. The appropriate excitations 
of that model are kinks, antikinks, and breathers. In the presence of 
an external voltage, the pseudoenergies of the kinks, antikinks 
(both with charge $e$), and breathers (with no charge) are affected by 
the bias $V$. At $\nu=1/3$, there is just one kink, one antikink, and 
one breather. The effect of the impurity is the scattering of these 
quasiparticles \footnote{The quasiparticles of the appropriate basis of
the boundary sine--Gordon model should not be confused with the Laughlin
QPS of the original problem. The relation between these two kinds of 
quasiparticles is nontrivial.}. 
Importantly, the quasiparticle basis is chosen in such 
a way that they scatter one by one (without particle production) 
on the boundary. This allows us to do a Landauer--B{\"u}ttiker--like 
transport theory for these quasiparticles. Although, there are just 
three types of quasiparticles at $\nu=1/3$, the calculation of 
the current noise is tedious and we briefly sketch it now. For
a more detailed description we refer to the original literature
\cite{fendl95,fendl96}. In the following, we set $h \equiv 1$ and only
restore it in the final results. In the field theoretic solution of the
CLL model, a system on a circle of large length $L$ is studied at
temperature $T$. Each quasiparticle has the energy $e_i$ ($i=+$ for the
kink, $i=-$ for the antikink, and $i=b$ for the breather); 
in a gapless theory $e_i= \pm p_i \propto e^\theta$. The level
density $n_i(\theta)$ ($\theta$ is the rapidity) and the filling
fraction $f_i(\theta)$ are introduced, so that $L n_i(\theta)f_i(\theta)$
is the number of allowed states for a quasiparticle of type $i$ in the
rapidity range $(\theta, \theta + d\theta)$. Therefore, the
density of occupied states per unit length is given by 
$P_i(\theta) = n_i(\theta)f_i(\theta)$. Requiring that the wave functions 
be periodic in space, this yields the Bethe equations
\begin{equation} \label{TBAeq}
n_i(\theta) = \frac{dp_i}{d\theta} + \sum_j \int d\theta' \Phi_{ij} 
(\theta - \theta') P_j(\theta) \; ,
\end{equation}
where the kernel $\Phi_{ij} (\theta - \theta')$ is related to the bulk
scattering matrix of the quasiparticles \cite{fendl95}. Furthermore, 
pseudoenergies $\epsilon_i$ are defined by
\begin{equation}
f_i = \frac{P_i}{n_i} \equiv \frac{1}{1+e^{\epsilon_i-\mu_i}} \; , 
\end{equation}
where $\epsilon_i$ and the chemical potentials $\mu_i$ are scaled by the
temperature to make them dimensionless. In our case $\mu_+ = eV/2k_B T$, 
$\mu_- = -eV/2k_B T$, and $\mu_b = 0$. In order to determine the 
equilibrium
values of the pseudoenergies, one has to find the configuration which
minimizes the free energy. This procedure is known as the
thermodynamic Bethe ansatz. After all we are able to calculate the current
noise defined by Eq.~(\ref{def_noise}) in this formalism exactly. The
final result is given by
\begin{eqnarray} \label{exactsol}
S &=& L \int \int d \theta d \theta' D(\theta,\theta') {\cal T} (\theta)
{\cal T} (\theta') \\ \nonumber
&+& \int d\theta \Bigl\{ \Bigl[ \overline{P}_+(\theta) 
\left( 1-\overline{f}_-(\theta) \right)
+ \overline{P}_-(\theta) \left( 1- \overline{f}_+(\theta) \right) \Bigr] 
\nonumber \\
&\times& {\cal T} (\theta) \left( 1- {\cal T} (\theta) \right) \Bigr\} 
\; , \nonumber
\end{eqnarray}
where $\overline{A}$ indicates thermal equilibrium of $A$ and the 
density--density correlator $D(\theta,\theta')$ reads
$D(\theta,\theta') \equiv \overline{\Delta (P_+ - P_-)(\theta) 
\Delta (P_+ - P_-)(\theta')}$.
In Eq.~(\ref{exactsol}), we have introduced the energy--dependent 
transmission coefficient ${\cal T} (\theta) \equiv 1/(1+ e^{2(g-1)(\theta-\theta_B)/g})$.
$\theta_B$ parametrizes the impurity strength and is related to
$\lambda$ in Eq.~(\ref{HB}) through $k_B T_B \equiv \frac{M}{2} e^{\theta_B} \propto
\lambda^{1/(1-g)}$ ($M$ is an arbitrary energy scale that cancels
out of all physical results) \cite{fendl95}.

Let us now turn back to the relevance of interactions in
tunnel experiments with Laughlin QPS; more specifically, we discuss
whether Eq.~(\ref{exactsol}) can be approximated by a 
formula like Eq.~(\ref{leso1}) with an appropriate adjustment of parameters.
In order to do so, we first find the best candidates to approximate
Eq.~(\ref{exactsol}). Eqs.~(\ref{kfwbl}) and 
(\ref{kfsbl}) suggest that the WBL and the SBL should be treated by
separate approximations. Moreover, the slope of the finite frequency
(non--equilibrium) noise around $\omega=0$ \cite{chamo99} indicates
that an identification of the dimensionless conductance $(1/G_\nu) dI/dV$
(with $G_\nu = \nu e^2/h$) as an energy--dependent transmission 
coefficient ${\cal T}(E) \equiv (1/G_\nu) dI/dV$ is plausible.
\begin{figure}
\includegraphics[angle=-90,width=7.0cm,keepaspectratio,clip]{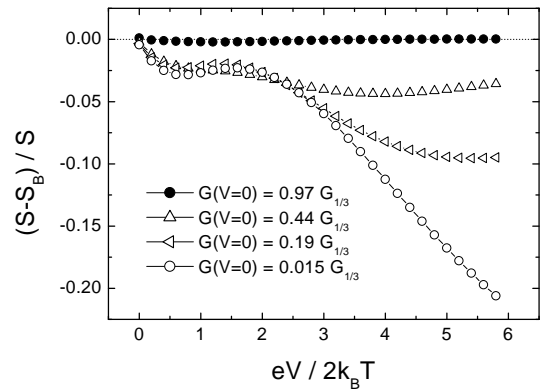}
\caption{The relative difference between the exact solution $S$ for $\nu=1/3$
and the approximation $S_B$  (under the choice $e^*=e/3$)
is plotted against $eV/k_B T$ for different values of the zero bias conductance 
$G$: $G=0.97 \ e^2/3h$, $G=0.44 \ e^2/3h$, $G=0.19 \ e^2/3h$, and 
$G=0.015 \ e^2/3h$. The non-monotone behavior of $(S-S_B)/S$ as a function
of $eV/k_B T$ for smaller values of $G$ is related to the fact that the noise
$S$ is highly nonlinear as a function of $eV/k_B T$ away from the WBL.}\label{fig_comp}
\end{figure}
An appropriate combination of Eqs.~(\ref{leso1}) and (\ref{kfwbl}) yields 
the tentative expression
\begin{equation}
S_B = 2e^* I_{\rm BS} {\cal T}(E) \coth\left(\frac{e^*
V}{2k_BT}\right)+4k_BTG_\nu {\cal T}^2(E) \, , \label{fitwbl}
\end{equation}
which is expected to satisfyingly approximate $S$ in the WBL.
We checked numerically for $eV/k_BT \in [0,12]$
that this is, indeed, the case. A comparison between Eqs.~(\ref{exactsol}) and
(\ref{fitwbl}) (under the choice $e^*=e\nu$ for $\nu=1/3$) at different values 
of the zero bias conductance, i.~e. different values of $\theta_B$, is illustrated in 
FIG.~\ref{fig_comp}. It is clearly visible that $S_B$ agrees very well
with $S$ over the whole range of $eV/k_BT$ in the WBL, see e.~g. the curve
for $G=0.97 \ e^2/3h$. Thus, there {\it does} exist a meaningful 
window of parameters where the effect of interactions can be accounted for 
by adjusting parameters in the non--interacting expression (\ref{fitwbl}). 

In the intermediate scattering regime, e.~g. for $G=0.44 \ e^2/3h$ in
FIG.~\ref{fig_comp}, the agreement between $S_B$ with $e^*=e/3$ and $S$ 
is still quite good (within a few percent). 
However, in the SBL, the case $G=0.015 \ e^2/3h$ in FIG.~\ref{fig_comp}, 
Eq.~(\ref{fitwbl}) with $e^* = e \nu$ diverges from 
the exact solution as $eV/k_B T$ grows. For that value of $G$, 
we have verified that the alternative expression
\begin{equation}
S_T = 2e I_T (1-{\cal T}(E)) \coth\left(\frac{e
V}{2k_BT}\right)+4k_BTG_0 {\cal T}^2(E) \label{fitsbl}
\end{equation}
almost coincides with the exact solution $S$ in the full
range $eV/k_BT \in [0,12]$. This is not very surprising since
Eq.~(\ref{fitsbl}) could have been guessed as the best approximation 
to the exact solution in the SBL by combining Eqs.~(\ref{leso1}) and 
(\ref{kfsbl}), i.~e., roughly, by exchanging Laughlin QPS with electrons.

\begin{figure}
\vspace{0.5cm}
\begin{center}
\epsfig{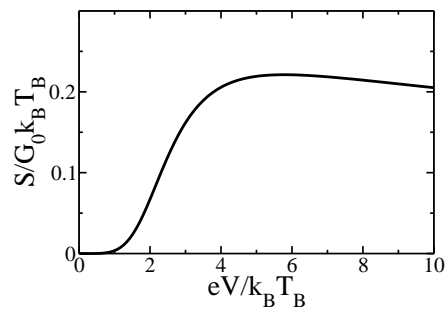}
\caption{\label{fig_v_vs_s} The shot noise (in units of $G_0 k_B T_B$)
is plotted against the applied voltage (in units of 
$k_B T_B/e$) for zero temperature 
and $g=1/3$. For small values of $eV/k_B T_B$, the noise first
increases as a powerlaw (in the SBL), then reaches a maximum, and, finally,
decreases as a powerlaw if $eV/k_B T_B$ becomes large (in the WBL). 
This curve describes well the measurements of current noise at 
very low temperatures reported in FIG.~2(b) of Ref.~\cite{chung03}.}
\end{center}
\end{figure}
This is of course not to say that the more sophisticated theory is useless.
There does exist as well a regime of parameters, where interaction 
effects are so pronounced that an analysis of the noise with 
interacting theories is unavoidable. As realized by the authors of
Ref.~\cite{chung03} this is the regime of low temperatures.
In that regime, the $dI/dV$ curves show strong nonlinearities and
the current noise cannot be fitted by non--interacting formulae 
anymore (see Figs.~1 and 2 in Ref.~\cite{chung03}). The
reason for this is apparent: For $V=0$ and $T=0$, the system is in the
SBL fixed point. If we now increase $V$ we leave the SBL and drive it
into the WBL. Thus, neither of the two approximations (\ref{fitwbl}) and
(\ref{fitsbl}) can be the good one in the whole range of the applied voltage. 
However, as shown in our FIG.~\ref{fig_v_vs_s}, the zero temperature shot 
noise for $g=1/3$ describes well the behavior of the shot noise 
as a function of voltage bias reported in FIG.~2(b) of Ref.~\cite{chung03}.

Finally, the validity of non--interacting formulae in some of the regimes of parameters, 
provided the effective charge is adjusted to $e^*$ or $e$, suggests considering 
the ``evolution of the charge of tunnelling objects'' with the 
scattering strength as measured in Ref.~\cite{griff00}. 
Let us stress that we do not believe that a charge other than $e^*=e\nu$
and $e$ has much significance: 
While, in the WBL, Laughlin QPS indeed tunnel independently, 
and, in the SBL, electrons do so, 
intermediate scattering regions can only be matched by a more complex 
description than independent particle tunnelling. However, one can always take  
the field theoretic calculation of the current noise, 
and extract a value of $e^*$ by fitting the 
exact solution to Eqs.~(\ref{fitwbl}) and (\ref{fitsbl}) for 
filling fraction $\nu=1/3$ with $e^*$ being a $\nu$--independent 
fitting parameter \footnote{In Eq.~(\ref{fitsbl}), $e$ has to be replaced 
by $e^*$ and $G_0$ by $G_\nu$ to do a self--consistent fit. 
Furthermore, $G_\nu$ has to be defined as $G_\nu \equiv (e^*/e) e^2/h$
during the fitting procedure.}.
\begin{figure}
\begin{center}
\epsfig{file=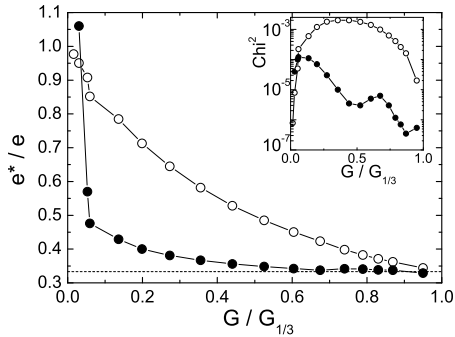,scale=0.63}
\caption{\label{fig_estar} Apparent evolution of fractional charge from the SBL to 
the WBL. The points are obtained by fitting the exact solution of the 
noise Eq.~(\ref{exactsol}) with Eq.~(\ref{fitwbl}) (black points) and 
Eq.~(\ref{fitsbl}) (white points) for 
$eV/k_B T \in [0,12]$. The inset shows the $\chi^2$ values of the 
two fits and $G$ is the zero bias conductance ($G_{1/3}=e^2/3h$). 
The dashed line corresponds to $e^*=e/3$.}
\end{center}
\end{figure}
In FIG.~\ref{fig_estar} the extracted values of $e^*$ are shown for 
comparison. Evidently, the fractional charge depends significantly 
on the choice of the heuristic formula used to extract it, especially
in the intermediate scattering regime. 
We see that the values of $e^*$, which have been 
obtained through Eq.~(\ref{fitsbl}), bear a lot of similarity to the 
measured values in FIG.~4 of Ref.~\cite{griff00}. This should be 
due to the fact that Eq.~(\ref{fitsbl}) is, although less complex, 
very similar to Eq.~(4) of Ref.~\cite{griff00}. From this analysis
we conjecture that the measured noise in Ref.~\cite{griff00} could have
been fitted quite well by the exact field theoretic solution 
for all values of zero bias transmission without any adjustment
of $e^*$.

In conclusion, we have clarified the subtle issue of the relevance of
interactions of Laughlin QPS at a constriction in a FQH bar by
analyzing the noise properties of such a system. For $\nu=1/3$, we have
quantitatively identified the regime of parameters, in which
the effect of interactions can be incorporated into non--interacting
expressions by an ``adjustment of parameters'', and, on the other hand,
the regime, in which an appropriate treatment of interactions
is crucial. Our findings are of high relevance for the interpretation of
measurements of shot noise in FQH systems.

Interesting discussions with Y.~C.~Chung, H.~Grabert, and 
M.~Heiblum are gratefully acknowledged. Financial support has been 
provided by the EU (BT) and the Humboldt Foundation (HS).


\begin{thebibliography}{10}

\bibitem{laugh83}
{R.~B.~Laughlin}, Phys.~Rev.~Lett. {\bf {50}},  1395  (1983).

\bibitem{wen90}
{X.~G.~Wen}, Phys.~Rev.~Lett. {\bf {64}},  2206  (1990); Phys.~Rev.~B 
{\bf {41}},  12838  (1990).

\bibitem{depic97}
{R.~de-Picciotto {\em et al.}}, Nature (London) 
{\bf {389}},  162  (1997).

\bibitem{samin97}
{L.~Saminadayar, D.~C.~Glattli, Y.~Jin, and B.~Etienne}, 
Phys.~Rev.~Lett. {\bf {79}},  2526  (1997).

\bibitem{glatt00} D.~C.~Glattli {\em et al.}, Physica E {\bf 6}, 22 
(2000).

\bibitem{griff00}
{T.~G.~Griffiths {\em et al.}}, Phys.~Rev.~Lett. {\bf {85}},  3918  
(2000).

\bibitem{chung03}
Y.~C.~Chung {\em et al.}, Phys.~Rev.~B {\bf 67}, 201104(R) (2003).

\bibitem{rosen01}
{B.~Rosenow and B.~I.~Halperin}, Phys.~Rev.~Lett. {\bf {88}},  096404  
(2002).

\bibitem{fishe97}
M.~P.~A.~Fisher and L.~I.~Glazman in {\it Mesoscopic Electron
Transport}, Vol. 345 of NATO ASI, edited by L.~Kouwenhoven,
G.~Sch{\"o}n, and L.~Sohn (Kluwer, Dordrecht, 1997).

\bibitem{lesov89}
{G.~B.~Lesovik}, JETP Lett. {\bf {49}},  592  (1989); 
{M.~B{\"u}ttiker}, Phys.~Rev.~Lett. {\bf {65}},  2901  (1990);
{Th.~Martin and R.~Landauer}, Phys.~Rev.~B {\bf {45}},  1742  (1992).

\bibitem{kane94}
{C.~L.~Kane and M.~P.~A.~Fisher}, Phys.~Rev.~Lett. {\bf {72}},  724  
(1994).

\bibitem{fendl95}
{P.~Fendley, A.~W.~W.~Ludwig, and H.~Saleur}, Phys.~Rev.~Lett. 
{\bf {74}}, 3005  (1995); {\bf {75}}, 2196  (1995); 
Phys.~Rev.~B {\bf {52}}, 8934  (1995).

\bibitem{fendl96}
{P.~Fendley and H.~Saleur}, Phys.~Rev.~B {\bf {54}},  10845  (1996).

\bibitem{chamo99}
{C.~Chamon and D.~E.~Freed}, Phys.~Rev.~B {\bf {60}},  1842  (1999).

\end{thebibliography}
\end{document}